\documentclass[12pt]{article}
\usepackage{amsfonts}
\usepackage[dvips]{graphicx}
\begin{document}
\title{\bf{How not to share a set of secrets}}
\author{K. R. Sahasranand\small$^{1}$ \and Nithin Nagaraj\small$^{1}$ \and Rajan S.\small$^{2}$ \and \small $^{1}$Department of Electronics and Communication Engineering,\\ \small $^{2}$Department of Mathematics, \\\small Amrita Vishwa Vidyapeetham, Amritapuri Campus, \\ \small Kollam-690525, Kerala, India.\\{\bf \small Email:~sanandkr@gmail.com, nithin@am.amrita.edu, rajans@am.amrita.edu}}
\date{}
\maketitle \abstract \noindent This note analyses one of the existing space efficient secret sharing schemes and suggests vulnerabilities in its design. We observe that the said algorithm fails for certain choices of the set of secrets and there is no reason for preferring this particular scheme over alternative schemes. The paper also elaborates the adoption of a scheme proposed by Hugo Krawczyk as an extension of Shamir's scheme, for a set of secrets. Such an implementation is space optimal and works for all choices of secrets. We also propose two new methods of attack which are valid under certain assumptions and observe that it is the elimination of random values that facilitates these kinds of attacks.\\

\noindent {\bf Keywords}: cryptography, secret sharing, set of secrets, space optimal.

\section{Introduction}

At times, we come across such situations wherein we want to share a secret among a set of people in such a way that if more than a particular number of people from that set come together, the secret could be reconstructed. However, any number of people less than that particular number, albeit from the same set of people, should not be able to learn anything about the secret. Such a scheme is called a threshold secret sharing scheme. If the size of the whole set of people is $n$ and the threshold is $k$, we call that scheme a $(k, n)$ threshold secret sharing scheme. In such a scheme, $n$ shares are generated and distributed among the people, any $k$ of them enough for reconstruction of the original secret while any $k-1$ or less will not be able to recover the secret. We might also come across cases where we have a set of secrets rather than a single piece of secret to be shared among a set of people. This set of secrets contains elements which may or may not be distinct.

In a ($k$, $n$) secret sharing scheme by Shamir \cite{shamir}, there results an $n$-fold increase in the total storage requirement. Further, in cases where $k$ secrets have to be shared among $n$ individuals for a ($k$, $n$) scheme, the storage requirement explodes to $k$.$n$ times the original. One of the already proposed schemes \cite{abhi} which claims to be space efficient, however, was found to compromise usability for efficiency, without really achieving efficiency. In this paper, we describe the vulnerabilities in its design and offer a better solution which overcomes these drawbacks.

\section{Review of existing scheme}

\noindent The scheme in \cite{abhi} works in the following manner:\\

\indent Consider a polynomial function $y = q(x)$ of degree $k-1$, from $\mathbb{Z}_p$ to $\mathbb{Z}_p$. The secrets $s_0, s_1, \ldots, s_{k-1}$ are treated as $y$ values with $x=0,1,\ldots,k-1$. Then, they are interpolated to form a $k-1$ degree polynomial and the value of $q(x)$ at $n$ different $x$ are calculated, where \mbox{$x \neq 0,1,\ldots,k-1$.} Each of the $n$ values thus obtained, along with the corresponding $x$ value is a share. Note that the number of secrets, $k$ is the same as the threshold value $k$ in this particular $(k, n)$ scheme.\\

\noindent A number of problems identified with this scheme follows:

\begin{enumerate}
\item It shares $k$ secrets among $n$ people such that any $k$ of them can reconstruct the secret. Thus the threshold $k$ has to be chosen to be equal to the number of secrets to be shared. Thus, for sharing $k$ secrets (for which the order of secrets matters as well as for which it doesn't) in an ($m$, $n$) scheme, where $m$ is the threshold, $m > k$, the scheme is insufficient. Clearly, it is undesirable that the number of secrets dictates the threshold to be used.\\
\label{item:prob_thresh}

\item The scheme cannot be used to implement a ($k$, $n$) scheme when the $k$ secrets to be distributed are inherently generated from a polynomial of order less than $k-1$. This may be demonstrated using an example.
 
Let $s_0$, $s_1$, $s_2$ and $s_3$ be four secrets in a finite field $\mathbb{Z}_p$ to be shared in a (4, $n$) scheme. If $s_0$ = 2, $s_1$ = 6, $s_2$ = 12, $s_3$ = 20, and $p$ = 31, the scheme expects a cubic polynomial to be the result of interpolation. However, the polynomial resulting out of interpolation happens to be $x^2 + 3x + 2$, which is quadratic. If the shares are generated from this polynomial, it requires only $3$ people for reconstruction; thereby a ($4$, $n$) scheme could not be implemented.

One solution to this problem appears to be changing the order of secrets or changing the indices $i$'s of $s_i$. However it costs a round of interpolation because it is impossible to guess such a relation between the seemingly {\it innocent} secrets without interpolating. Besides, changing the order may not always be feasible (for example, when these secrets are pieces of a larger secret) and changing the indices may not always work.\label{item:prob_dep}\\

\item The scheme does not work if all of the secrets to be shared are the same. This case is quite possible; when the $k$ secrets are part of a large secret like it is mentioned in \cite{abhi}. But, for the scheme to work, there should be at least one $s_i \ne s_j$ among all $s_i , s_j$ to be shared. This is by definition of interpolation. For successful interpolation over a finite field, we need at least two distinct values.

For example, if we want to share $3$ secrets, all of them equal to $2$, taken from a finite field $\mathbb{Z}_p$, $p = 3$. Then, $s_0 = 2$, $s_1 = 2$, $s_2 = 2$ and $p = 3$. Interpolation is performed as follows:\\

$q_0(x) = \frac{x-1}{0-1}.\frac{x-2}{0-2} = 2x^2 + 1$.\\

$q_1(x) = \frac{x-0}{1-0}.\frac{x-2}{1-2} = 2x^2 + 2x$.\\

$q_2(x) = \frac{x-0}{2-0}.\frac{x-1}{2-1} = 2x^2 + x$.\\

Now, $q(x) = s_0.q_0(x) + s_1.q_1(x) + s_2.q_2(x)$.\\

$\Rightarrow q(x) = 2.(q_0(x) + q_1(x) + q_2(x))$.\\

$\Rightarrow q(x) = 2$.\\

Thus, we end up with a constant polynomial. The scheme needs a quadratic polynomial for implementing a $(3, n)$ scheme and hence does not work in such cases. It could be easily seen that $k$ secrets could be chosen all of them being the same, from a field $\mathbb{Z}_p$ in $p$ ways. There is a total of $p^k$ ways by which any $k$ secrets could be chosen from $\mathbb{Z}_p$ (the $k$ secrets may or may not be distinct). The method in \cite{abhi} does not work in any of the $p$ cases out of the $p^k$ cases possible.\\

\label{item:prob_repeat}

\item The percentage of cases for which the method in \cite{abhi} fails may be calculated as follows:
\noindent Here we have assumed that the order of secrets is important. i.e., a set of secrets, say, $\{0,1,1,2\}$ is different from $\{0,1,2,1\}$.  

In general, there are \mbox{$p^{k-1}$} cases in which the $k$ shares are generated from polynomials of order less than $k-1$. This includes the cases in which all the secrets are the same. The count is obtained by the following argument:

We need to find the number of such instances where the secrets $s_0, s_1, \ldots, s_{k-1}$ are generated from polynomials of order strictly less than $k-1$. Let the polynomial be named $f(x)$. Note that all operations are performed modulo $p$. Thus,

\begin{center}
$f(x) = a_{k-1}x^{k-1} + \ldots + {a_2}x^2 + {a_1}x + a_0$; \space \space $a_{k-1} = 0$.
\end{center}

Now, the secrets are the values of $f(x)$ at different values of $x$, $0 \le x \le k-1$.
\begin{eqnarray*}
& s_0 = f(0) = 0 + \ldots + 0 + 0 + a_0,\\
& s_1 = f(1) = a_{k-1} + \ldots + {a_2} + {a_1} + a_0,\\
& \vdots \\
& s_{k-1} = f(k-1) = a_{k-1}{(k-1)}^{k-1} + \ldots + {a_2}{(k-1)}^2 + {a_1}{(k-1)} + a_0.\\
\end{eqnarray*}

In matrix form, this looks like,

\begin{displaymath}
\left[ \begin{array}{cccc}
0 & 0 & \ldots & 1\\
1 & 1 & \ldots & 1\\
2^{k-1} & \ldots & \ldots & 1\\
\vdots & \vdots & \ddots & \vdots\\
(k-1)^{k-1} & \ldots & \ldots & 1
\end{array} \right]\cdot
\left[ \begin{array}{c}
a_{k-1}\\
a_{k-2} \\
\vdots \\
a_1\\
a_0
\end{array} \right]
=
\left[ \begin{array}{c}
s_0\\
s_1 \\
\vdots \\
s_{k-2}\\
s_{k-1}
\end{array} \right].
\end{displaymath}\\

Let us denote the above matrices respectively as $M$, $A$ and $B$.
Thus, $M \cdot A = B$.\\
Now, the matrix $M$ is the Vandermonde matrix of order $k$ and is essentially invertible \cite{vander}. Furthermore, even under modulo $p$, this is true since the determinant of $M$ is not divisible by $p$ \cite{wolf}. Therefore, we can write, $A = M^{-1} \cdot B$.\\

\begin{displaymath}
\left[ \begin{array}{c}
a_{k-1}\\
a_{k-2} \\
\vdots \\
a_1\\
a_0
\end{array} \right]
= M^{-1}\cdot
\left[ \begin{array}{c}
s_0\\
s_1 \\
\vdots \\
s_{k-2}\\
s_{k-1}
\end{array} \right].
\end{displaymath}

If $m_1, m_2, \ldots, m_k$ are the elements of first row of $M^{-1}$, then, since $a_{k-1} = 0$, we have,

\begin{center}
$m_0s_0 + m_1s_1 + \ldots + m_{k-1}s_{k-1} = 0$.
\end{center}

Here, we have $(k-1)$ ways to choose the secrets $s_0, \ldots, s_{k-2}$. From the above equation, the $k$-th secret, $s_{k-1}$ has to be the negative of the sum of the rest of the secrets mod $p$. Thus, the resulting number of choices for the set of secrets is ${p}^{k-1}$, considering that the same set of numbers taken in a different order is treated as a different set of secrets.\\

\noindent This count may be obtained by an alternative approach as well (We thank the author of \cite{abhi} for this argument).

The $k$ secrets being generated from a polynomial of order less than $k-1$ occurs in the following manner. The polynomial constructed out of the remaining $k-1$ secrets passes through the value of the $k$-th secret $s_{k-1}$ at $x = k-1$. i.e., the set we consider for interpolation is an oversampling of a lower degree polynomial. \\
Now, having chosen $k-1$ secrets from a finite field $\mathbb{Z}_p$, the $k$-th secret $s_{k-1}$ could be just one of the possible $p$ choices. This is because $s_{k-1}$ is treated as the $y$ value at $x = k-1$. The polynomial constructed out of the remaining $k-1$ secrets passes through exactly one point corresponding to $(k-1, s_{k-1})$. However, the $k-1$ secrets could be chosen from $\mathbb{Z}_p$ in ${p}^{k-1}$ ways. This value multiplied with the number of choices for the $k$-th secret $s_{k-1}$ (which is equal to $1$) gives ${p}^{k-1}$ cases in which the method fails.\\

Thus, there are ${p}^{k-1}$ choices of secrets out of the $p^k$ choices possible which yield polynomials of degree less than $k-1$ upon interpolation, thereby leaving us unable to implement a $(k, n)$ scheme.

\begin{equation}\label{eq:fails}
\emph{Percentage of failure} = \frac{p^{k-1}}{p^{k}}\cdot100 = 100/p.
\end{equation}
\label{item:prob_per}

\item It is mentioned in \cite{abhi} that the primes chosen are of the order of $1024$ bits. Then, even for small $k$, there is a large number of choices of secrets which could not be implemented using the scheme. For large values of $p$, the failure percentage is small as indicated by the formula. However, since the failure rate is not zero, it is absolutely necessary to verify whether for a particular set of secrets the method works or not.\\
For example, for $k=3$ and $p$ of the order of $1024$ bits, number of cases for which the method fails, $f=$ ${p}^{k-1}$. $f$ is of the order of  $2^{1024}.2^{1024} = 2^{2048}$. The number of cases for which the scheme in \cite{abhi} fails is of the order of $2048$ bits for a simple $k=3$. i.e., there are about $2^{2048}$ choices of a set of $3$ secrets for which the scheme does not work (The number $2^{2048}$ is nearly $600$ digits long) although this is only a very small percentage of the possible set of secrets. The problem lies in the fact that it is impossible to know beforehand whether or not the method could be used for a particular $(k, n)$ scheme for a given set of secrets.

\end{enumerate}

\section{Definitions}
Borrowing from \cite{rec}, blow up factor is defined in various cases as follows,\\

\noindent {\bf Definition 1}. {\it Blow-up factor (secret sharing)}\\
\\ \indent $= \frac{\emph{Total size of shares}}{\emph{Total size of secrets encoded by the shares}}$\\

$=\frac{\emph{Number of shares . Size of a share}}{\emph{Total size of original secrets}}$\\

\noindent {\bf Definition 2}.  {\it Blow-up factor (conventional secret sharing)}\\

\indent $ = \frac{\emph{Number of shares . Size of a share}}{\emph{Total size of original secrets}}$\\

$ = \frac{\emph{n.d}}{\emph{d}}$\\

\noindent {\bf Definition 3}.  {\it Blow-up factor (Space optimal secret sharing)}\\

\noindent Such a scheme is one with a blow up factor of $n / k$,
where $2 \le k \le n$. This is because for a space optimal secret sharing scheme,\\

$\frac{\emph{Number of shares . Size of a share}}{\emph{Total size of original secrets}}$\\

$=\frac{{n.d}}{{k.d}}$\\

$=\frac{\emph{n}}{\emph{k}}$

\section{A better solution}
\noindent A scheme with the following properties is desirable:

\begin{itemize}
\item A space optimal secret sharing algorithm for $k$ secrets in an ($m$, $n$) scheme, $m\geq k$. i.e., it does not demand that the threshold be chosen to be equal to the number of secrets. 
\item A scheme that allows for repetition of secrets and works when the order of secrets matters as well as when it doesn't.
\item An implementation friendly scheme where the need for random values is completely eliminated or minimized in certain cases.
\end{itemize}

\subsection*{Krawczyk's algorithm}
\noindent Krawczyk outlines an algorithm in \cite{kraw} for information dispersal. The same method when used for sharing a set of secrets does not have any of the drawbacks the method in \cite{abhi} has and is space optimal as well. Let us first consider the case of $k$ secrets to be shared among $n$ people in a $(k, n)$ scheme.\\

\noindent {\it Algorithm:}

\begin{enumerate}

\item Construct a polynomial $q(x)$ in such a way that the secrets are the coefficients of powers of $x$. If $s_0$, $s_1$, $s_2$,\ldots, $s_{k-1}$ are the $k$ secrets, then the polynomial to be constructed is:

\begin{center}
$q(x) = s_0 + {s_1}x + {s_2}x^2 + \ldots + s_{k-1}x^{k-1}$.
\end{center}

Here, we have to ensure that the coefficient of the highest power of $x$ is non-zero. i.e., $s_{k-1} \ne 0$
\item Compute the values of $q(x)$ at $n$ different values of $x$, say $x_0$, $x_1$, \ldots, $x_{n-1}$  and distribute them as shares $(x_0, b_0)$, $(x_1, b_1)$, \ldots, $(x_{n-1}, b_{n-1})$ where $b_i = q(x_i), 0 \le i \le n-1$.\label{algo:compute}

\item Reconstruction of secrets could be carried out using any $k$ of the values generated in step~\ref{algo:compute} above, by solving for $s_i$s or obtaining $q(x)$ through interpolation.

\end{enumerate}

The aforementioned scheme works for all possible set of secrets except in the trivial case of all secrets being equal to $0$. Through this scheme, we have attained the optimal $(n / k)$ blow up factor in total storage size. i.e., storing $k$ secrets, each of length $d$, demands only a storage space of $d.n$. Besides, it eliminates the need for random coefficients for powers of $x$ as secrets themselves are assumed here to be truly random. If at all the secrets are not random but follow a certain distribution, they could be randomised suitably (using a hash function) for serving our purpose.

Further, this scheme could be extended to implement a $(m, n)$ scheme for sharing $k$ secrets, where $m \ge k$. This could be achieved by involving random coefficients for powers, $j$ of $x$, where $k \le j \le m$.

Such a design is free of the constraint of the algorithm in \cite{rec}, where only a $(k, n)$ scheme could be implemented with $k-1$ secrets. Besides, the method in \cite{rec} involves $k-1$ interpolations for reconstruction of secret since it uses a recursive scheme. After each interpolation, a polynomial is obtained with the free term corresponding to one of the secrets. The proposed method, on the other hand, uses just one interpolation. Thus it has improved space efficiency and is faster in share generation as well as reconstruction.

In this method, since the secrets themselves are assumed to be \emph{truly random} by nature, they are equivalent to the random coefficients in \cite{shamir}. Since, Shamir's scheme is unconditionally secure, so is this scheme.

\subsection*{Sharing a very large secret}

\noindent The algorithm proposed above could be extended to conventional schemes like \cite{shamir} to eliminate the need for random values. It provides computational security in cases where the length of the secret is very big and causes storage and computation inconvenience. The algorithm in \cite{kraw} was originally meant to serve this purpose. It is implemented in the following manner:

A large secret of length, say 20000 bits, has to be shared among $n$ people such that any 10 of them can reconstruct the secret. Then, instead of using random coefficients for the 9 powers of $x$, namely, $x$, $x^2$, \ldots, $x^9$, the 20000 bit secret is split into 10 pieces 2000 bits each. These are then used as the coefficients of the powers of $x$ from 0($x^0$, free term) to 9($x^9$). The possible values of the secret now span $\mathbb{Z}_p$, where $p$ is of the order of 2000 bits instead of 20000 bits. The advantages of this kind of implementation are:
\begin{itemize}
\item Primes of the order of 2000 bits are easier to find and use compared to 20000 bits.
\item Share generation and interpolation for reconstruction are faster.
\item It saves storage space.
\end{itemize}
However, note that the search space for a brute force search has now reduced.

\section{Curious cases of insecurity}

\noindent In this section we consider a couple of cases wherein the opponent (Eve) is assumed to possess partial information about the secrets. The proposed scheme as well as the scheme in \cite{abhi} are seen to fail under these assumptions. The failure could be attributed to them not employing random values.\\

\begin{bf} \noindent Case 1: \end{bf}\\

Here we elaborate a case in which the modified Krawczyk's scheme we proposed fails. The scheme was shown to be as much secure as Shamir's scheme under the assumption that the secrets themselves are truly random; i.e., they come from a uniform distribution over $\mathbb{Z}_p$. However, this may not really be the case. Suppose the set of secrets is known to come from a set whose members are much smaller than $p$ that modular arithmetic does not come into play at all.

Consider a $(k,n)$ scheme. Let the the polynomial be $q(x)$ and prime $p$ we use be large (compared to secrets). Assume that Eve happens to learn that the $k$ secrets are all less than some $r$, $r << p$.\\

$q(x) = a_{k-1}x^{k-1} + \ldots + {a_2}x^2 + {a_1}x + a_0$\\

\noindent Now, Eve being a share-holder as well, gets some arbitrary share $(u, q(u))$ at $x = u$. Knowing that the secrets, i.e., the coefficients $a_{k-1}, \ldots, a_2, a_1, a_0$ are all less than or equal to $r$, Eve infers that:\\

\begin{bf} If $q(u)$ is a multiple of $u$, then the secret $a_0$ is a multiple of $u$. The converse is also true. \end{bf}\\

\noindent Let us examine why this is so. $q(u)$ is the value of $q(x)$ sampled at $x = u$. Thus, $q(u) = a_{k-1}u^{k-1} + \ldots + {a_2}u^2 + {a_1}u + a_0$. All the terms on the $RHS$ (excluding $a_0$) being multiples of powers of $u$ are obviously multiples of $u$. If $LHS$ namely $q(u)$, is a multiple of $u$, it implies $a_0$ is a multiple of $u$ as well and vice versa. For example,

\noindent Suppose $q(x) = 4x^{3} + 3x^2 + 2x + 15$ and $p = 999961$. Note that $a_0 = 15$.

$q(3) = 108 + 27 + 6 + 15$ mod $999961$.

$q(3) = 156$ mod $999961$. 

$q(3) = 156$ is a multiple of $3$ $\Leftrightarrow$ $a_0$ is a multiple of $3$.

\noindent Suppose $q(x) = 4x^{3} + 3x^2 + 2x + 14$ and $p = 999961$. Note that $a_0 = 14$.

$q(3) = 108 + 27 + 6 + 14$ mod $999961$.

$q(3) = 155$ mod $999961$. 

$q(3) = 155$ is NOT a multiple of $3$ $\Leftrightarrow$ $a_0$ is NOT a multiple of $3$.\\

\noindent Thus, in cases where the opponent knows that all the secrets are much smaller than the publicly known $p$ and are all less than some $r$, she can safely assume that the statement above is true. If $q(u)$ is a multiple of $u$, the search space (number of choices for the secret) for $a_0$ is reduced from $r$ to the set of multiples of $u$ less than or equal to $r$. i.e., from $r$ to $\lfloor r/u \rfloor + 1$. If $q(u)$ is NOT a multiple of $u$, the search space for $a_0$ is reduced to $r - \lfloor r/u \rfloor -1$.

Such a problem does not occur with conventional Shamir's scheme since in that one, even if the secret is known to be less than or equal to $r$, the coefficients used are random (and distributed over the whole $\mathbb{Z}_p$) and hence the divisibility argument does not hold. Thus, even when the opponent knows that the secret is less than or equal to $r$, the brute force search space is still $r$ and not reduced.\\

\begin{bf} \noindent Case 2: \end{bf}\\

This case establishes the failure of the scheme we proposed as well as the scheme in \cite{abhi}. Suppose a set of $k$ secrets from $\mathbb{Z}_p$ has been shared using the scheme proposed in \cite{abhi} among a set of people $R$, the polynomial be $q(x)$. Another set of secrets from $\mathbb{Z}_p$, which are all some $d$ times the secrets in the first set are shared among a set of people $S$, the polynomial be $r(x)$. Let Eve be one of the share holders in $R$ but not $S$ and she gets some arbitrary share $(u, q(u))$. She can immediately infer that $r(u) = d.q(u)$ (mod $p$).

For example, suppose we want to share $3$ secrets taken from a finite field $\mathbb{Z}_p$ among a set of people $R$. Let $s_0, s_1, s_2$ be the secrets. Interpolation is performed as follows:\\

$q(x) = s_0.\frac{(x-1)(x-2)}{(0-1)(0-2)} + s_1.\frac{(x-0)(x-2)}{(1-0)(1-2)} + s_2.\frac{(x-0)(x-1)}{(2-0)(2-1)}$ (mod $p$).\\

\noindent If the secrets to be shared among the second set $S$ are $d.s_0, d.s_1$ and $d.s_2$ from $\mathbb{Z}_p$, then the polynomial is as:\\

$r(x) = d.s_0.\frac{(x-1)(x-2)}{(0-1)(0-2)} + d.s_1.\frac{(x-0)(x-2)}{(1-0)(1-2)} + d.s_2.\frac{(x-0)(x-1)}{(2-0)(2-1)}$  (mod $p$).\\

$\Rightarrow r(x) = d.q(x)$ (mod $p$).\\

\noindent Thus, being in possession of $q(u)$, Eve can participate in the reconstruction of the secret shared among the people in the second set $S$ since $r(u) = d.q(u)$ (mod $p$). Although Eve is authorised to participate only in the reconstruction of secret shared among the first set $R$, she can use the partial information (that the secrets are multiples of each other) to her advantage due to the flaw in the scheme. Note that such a problem does not arise with Shamir's scheme (which shares a single secret) since it employs random coefficients and hence the polynomial is not completely determined by the secret. The shares generated even for related secrets could be totally unrelated in Shamir's scheme.\\

The problems pointed out in the cases above are due to the fact that both the schemes do away with random numbers. We observe that in special cases as these, to share $k$ secrets it is advisable to use Shamir's scheme $k$ times to generate $k$ shares. Such an implementation, although not space efficient is perfectly secure.

\section{Conclusion}

\noindent
We have pointed out the vulnerabilities of an existing secret sharing scheme. We have demonstrated clearly that there is no reason to choose a scheme which fails for a certain percentage of cases when there is an obvious better alternative solution which works for all possible cases. We have also proposed a couple of new modes of attack under certain assumptions and have thus assessed the role of random values in strengthening a scheme. The modes of attack proposed could prove to be useful elsewhere as well.

\end{document}